%% file: main.tex
\documentclass[11pt,a4paper]{article}

\usepackage{jheppub}

\usepackage{amsmath,amssymb}
\usepackage{enumerate}
\usepackage{amsfonts}

\usepackage{subfigure}

\usepackage{allpurpose}

\usepackage{tikz}

\usepackage[marginal,perpage,symbol]{footmisc}

\usepackage{slashed} %

\graphicspath{{./figures/}}

\everymath{\displaystyle}
\makeatletter
\gdef\@fpheader{}
\makeatother

\usepackage[nottoc,notlof,notlot]{tocbibind}

\usepackage{indentfirst}

\begin{document}
\title{Fermionic Love number of higher-dimensional Reissner-Nordstr\"om black holes}
\date{\today}

\author[a,b,*]{Xiankai Pang,\note{The corresonding author.}}
\emailAdd{xkpang@cwnu.edu.cn}

\affiliation[a]{School of Physics and Astronomy, China West Normal University, Nanchong 637009, China}
\affiliation[b]{School of Physics and Astronomy, Beijing Normal University, Beijing 100875, China}

\author[c,d]{Yu Tian,}
\affiliation[c]{School of Physical Sciences, University of Chinese Academy of Sciences, Beijing 100049, China}
\affiliation[d]{International of Theoretical Physics, Chinese Academy of Science, Beijing 100190, China}
\emailAdd{ytian@ucas.ac.cn}

\author[b,e]{Hongbao Zhang}
\emailAdd{hongbaozhang@bnu.edu.cn}

\affiliation[e]{Key Laboratory of Multiscale Spin Physics, Ministry of Education, Beijing Normal University, Beijing 100875, China}

\author[a]{and Qingquan Jiang}
\emailAdd{qqjiangphys@yeah.net}

\abstract{
In this paper, we generalize our previous work on the fermionic tidal Love numbers (TLNs) to higher-dimensional Reissner-Nordstr\"om black holes.
The massless Dirac equation is solved in $D$-dimensional spacetime using ingoing Eddington coordinates and regular tetrads. After identifying the regular solution branch, we extract the fermionic TLNs from its asymptotic behavior at infinity. The resulting TLNs exhibit a rich dimension-dependent structure that generalizes the four-dimensional case. Unlike bosonic TLNs, which vanish for certain values of the total angular momentum $l$ in dimensions $D>4$, fermionic TLNs remain non-zero for all $l$ and $D \geq 4$, except for extremal black holes. Moreover, the $l$-dependence weakens as $D$ increases, disappearing entirely in the infinite-dimensional limit. These results provide new insights into black hole responses to fermionic perturbations in higher-dimensional spacetimes.
}


\maketitle

\section{Introduction}
\label{sec:introduction}

Tidal Love numbers (TLNs) quantify the deformation of compact objects in external tidal fields~\cite{Love:1909yed,Binnington:2009bb,Bhatt:2023zsy,Rodriguez:2026iot,Chakraborty:2026qru}. They imprint characteristic signatures on the phase evolution of gravitational waves from binary inspirals, making them crucial in gravitational-wave astronomy~\cite{Chakravarti:2018vlt,LeTiec:2020spy}. Furthermore, TLNs provide a unique window into the internal structure of compact objects, distinguishing neutron stars from black holes and probing exotic states of matter~\cite{Flanagan:2007ix,Hinderer:2007mb,Damour:2009vw,Chatziioannou:2020pqz}.

In \acrfull{gr}, asymptotically flat black holes in four dimensions are particularly simple, their static \acrshort{tln}s with respect to bosonic perturbations (scalar, electromagnetic, and gravitational) vanish identically~\cite{Poisson:2014gka,Landry:2015zfa,Pani:2015hfa,LeTiec:2020bos,Chia:2020yla}. This vanishing is understood as a consequence of hidden ladder symmetries~\cite{Hui:2021vcv,Hui:2022vbh,BenAchour:2022uqo,Sharma:2024hlz,Rai:2024lho,Combaluzier-Szteinsznaider:2024sgb,Berens:2025lsh,DeLuca:2025zqr,Cvetic:2026wht,Ghosh:2026vig}. Non‑trivial \acrshort{tln}s can arise in several contexts. In particular, dynamical tides are generically non‑vanishing~\cite{Chakraborty:2023zed,Chakraborty:2025wvs,Combaluzier--Szteinsznaider:2025eoc,Solon:2026ubm,Saketh:2026trm}. Even in the static regime, non‑vanishing \acrshort{tln}s can emerge from modified gravity theories~\cite{Cardoso:2018ptl,Bhattacharyya:2025slf,Barbosa:2025uau}, quantum‑gravity corrections~\cite{Gurlebeck:2015xpa,Nair:2022xfm,Motaharfar:2025ihv,Motaharfar:2025typ,Motaharfar:2025izo,Liu:2025iby}, or spacetimes with a non‑zero cosmological constant~\cite{Franzin:2024cah,Nair:2024mya}. Within \acrshort{gr} itself, non-zero static \acrshort{tln}s become non‑zero in higher-dimensional spacetimes~\cite{Kol:2011vg,Hui:2020xxx,Rodriguez:2023xjd,Chakravarti:2018vlt,Chakravarti:2019aup,Cardoso:2019vof,Singha:2025xah} or for perturbations of electrically charged scalar fields~\cite{Ma:2024few,Pereniguez:2025jxq}. Most recently, it has been shown that black holes can also acquire non‑zero \acrshort{tln}s in response to static \emph{fermionic} perturbations~\cite{Chakraborty:2025zyb,Pang:2025myy}, opening a new frontier in the study of tidal deformability.

In this paper, we extend the nascent study of static fermionic \acrshort{tln}s to higher-dimensional spacetimes. Specifically, we present, for the first time to our knowledge, the static fermionic \acrshort{tln}s for higher-dimensional \acrfull{rn} black holes. We show that the radial equation for massless and static fermionic perturbations maintains a form similar to the $4$-dimensional case, but with a crucial dependence on the spacetime dimension $D$. This allows us to derive the static fermionic \acrshort{tln}s analytically. A key finding is that, in contrast to the bosonic case where static \acrshort{tln}s vanish for perturbations with certain values of the total angular momentum $l$~\cite{Kol:2011vg,Hui:2020xxx,Ma:2024few}, the static fermionic \acrshort{tln}s of \acrshort{rn} black holes are non-vanishing for all $l \geq 1/2$. Our results reveal a rich, dimension-dependent structure in fermionic tidal responses, offering new insights into the deformability of black holes in higher dimensions.

The paper is organized as follows. In Section~\ref{sec:diracequation}, we derive the reduced Dirac equation for a massive spinor in a $D$-dimensional spacetime. We separate the angular part using a conformal transformation and employ ingoing Eddington coordinates with an associated regular tetrad to enforce horizon regularity of the spinor. In Section~\ref{sec:Lovenumber}, we focus on massless, static perturbations and derive the corresponding radial equation. We then extract the static fermionic \acrshort{tln}s from the asymptotic expansion of the regular solution. We conclude in Section~\ref{sec:discussion} by discussing the implications of our findings and outlining potential future research directions.

Throughout this paper, we adopt the metric signature $(-+\cdots +)$.

\section{The Dirac equation for $D$-dimensional \acrshort{rn} black hole}
\label{sec:diracequation}

Under the signature $(-+\cdots +)$, the Dirac equation for an electrically neutral fermion in $D$-dimensional spacetime can be written as follows~\cite{Weinberg:1995mt,Lopez-Ortega:2009flo,Liu:2019rbq,Pang:2024tco}
\begin{IEEEeqnarray}{rCl}
	\left(\slashed{\nabla}-m\right)\psi&=& 0 , \label{eq:diraceqpsi}
\end{IEEEeqnarray} 
where $m$ is the mass of the Dirac spinor $\psi$, and $\slashed{\nabla}=\gamma^\mu\nabla_{\mu}=\gamma^{\mu}\left(\partial_{\mu}-\Gamma_{\mu}\right)$ for coordinate-dependent gamma matrices $\gamma^{\mu}=\gamma^a e_{a}^{~\mu}$ and spin connection $\Gamma_{\mu}$\footnote{We employ Greek letters (e.g. $\mu$) for spacetime indices and Latin letters (e.g. $a$) for internal Lorentz indices.}. 
The constant gamma matrices $\gamma^a$ satisfy the Dirac algebra\footnote{The explicit form of the gamma matrices in $D$-dimensional spacetime is provided in subsection~\ref{sec:gammamatrices}.}
\begin{IEEEeqnarray}{rCl}
	\{\gamma^a,\gamma^b\} &=& 2\eta^{ab} , \label{eq:gammamatricesetaab}
\end{IEEEeqnarray} 
where $\eta^{ab}$ denotes the $D$-dimensional Minkowski metric. The spacetime metric $g_{\mu\nu}$ and the tetrad fields $e_a^{~\mu}$ are related via~\cite{Wald:1984rg} 
\begin{IEEEeqnarray}{rCl}
	e_{a}^{~\mu}e_{b}^{~\nu}\eta^{ab} &=& g^{\mu\nu}. 
\end{IEEEeqnarray} 
The spin connection $\Gamma_{\mu}$ is defined as
\begin{IEEEeqnarray}{rCl}
	\Gamma_{\mu} &=& -\frac{\ii}{2}\Gamma_{abc}\Sigma^{ab}e^c_{~\mu} , 
\end{IEEEeqnarray} 
in terms of the Ricci rotation coefficients
\begin{IEEEeqnarray}{rCl}
	\Gamma_{abc} &=& (e_{a\mu})_{;\nu}e_b^{~\mu}e_c^{~\nu}= \bigl(\partial_{\nu}e_{a\mu}-\Gamma^{\lambda}_{\mu\nu}e_{a\lambda}\bigr) e_b^{~\mu}e_c^{~\nu} ,
\end{IEEEeqnarray} 
and the Lorentz group generators
\begin{IEEEeqnarray}{rCl}
	\Sigma^{ab} &=& \frac{\ii}{4}[\gamma^a,\gamma^b] . 
\end{IEEEeqnarray}

\subsection{The $D$-dimensional \acrshort{rn} metric}
In $D$-dimensional spacetime, the \acrshort{rn} black hole and the electromagnetic potential $A_{\mu}$ can be written using the ingoing Eddington coordinate $v$ as follows\footnote{The ingoing Eddington coordinate is defined as $v=t+r_*$ for $(t,r)$ with  
\begin{IEEEeqnarray*}{rCl}
		\dd s^2  &=& -f(r)\dd t^2 +\frac{1}{f(r)}\dd r^2 + r^2 \dd \Sigma_{D-2}^2,~A_{\mu}=  -\sqrt{\frac{D-2}{2(D-3)}}\frac{Q}{r^{D-3}}(\dd t)_{\mu} . 
\end{IEEEeqnarray*} 
And $r_*$ is the tortoise coordinate such that $\dd r_*=\frac{1}{f(r)}\dd r$. Note that we have written the electromagnetic potential $A_{\mu}$ for completeness, although it does not affect the field equations since our Dirac fermion is electrically neutral.} 
\begin{IEEEeqnarray}{rCl}
	\dd s^2  &=& -f(r)\dd v^2 +2\dd v \dd r + r^2 \dd \Sigma_{D-2}^2 , \label{eq:metriceddington} \\
	A_{\mu}&=&  -\sqrt{\frac{D-2}{2(D-3)}}\frac{Q}{r^{D-3}}(\dd v)_{\mu} + \sqrt{\frac{D-2}{2(D-3)}}\frac{Q}{r^{D-3}f(r)}(\dd r)_{\mu} .
\end{IEEEeqnarray} 
where the blackening factor is given by 
\begin{IEEEeqnarray}{rCl}
	f(r) &=& 1-\frac{M}{r^{D-1}} +\frac{Q^2}{r^{2(D-3)}}, \label{eq:blackeningfactor}
\end{IEEEeqnarray} 
and 
\begin{IEEEeqnarray}{rCl}
	\dd \Sigma_{D-2}^2&=& \left(\dd\phi^1\right)^2 +\Sigma_{i=2}^{D-2}\prod_{j=1}^{i-1}\sin^2\phi^j \left(\dd\phi^i\right)^2 
\end{IEEEeqnarray} 
is the line element of a $(D-2)$-dimensional unit sphere with the volume $\omega_{D-2}=\frac{2\pi^{\frac{D-1}{2}}}{\Gamma \left(\frac{D-1}{2}\right)}$. 
The ADM mass and the charge of the black hole are given by~\cite{Liu:2019rbq} 
\begin{IEEEeqnarray}{rCl}
	\mathcal{M} &=& \frac{D-2}{16\pi}\omega_{D-2}M , ~\mathcal{Q}=\frac{\sqrt{2(D-2)(D-3)}}{8\pi}\omega_{D-2}Q .
\end{IEEEeqnarray} 
As in the $4$-dimensional case, $f(r)=0$ still has two positive roots (for non-extremal black holes), corresponding to outer and inner horizons, respectively
\begin{IEEEeqnarray}{rCl}
	r_+ &=& \left(M+\sqrt{M^2-Q^2}\right)^{\frac{1}{D-3}} , ~r_- = \left(M-\sqrt{M^2-Q^2}\right)^{\frac{1}{D-3}} . \label{eq:rprmdef}
\end{IEEEeqnarray} 

It's customary to introduce the following conformal transformation in $D$-dimensional spacetime~\cite{Lopez-Ortega:2009flo}
\begin{align}
	\tilde{g}_{\mu\nu} &= \Omega^2 g_{\mu\nu} ,~\tilde{e}_a^{~\mu}=\Omega^{-1}e_{a}^{~\mu},~\tilde{\psi}=\Omega^{-\frac{D-1}{2}}\psi,~\tilde{m}=\Omega^{-1}m, ~\tilde{\gamma}^a=\gamma^a,  \label{eq:conformaltransformation}
\end{align} 
under which the Dirac equation can be rewritten as follows~\cite{Lopez-Ortega:2009flo,Liu:2019rbq} 
(see  appendix~\ref{sec:diracequationconformal} for a proof)
\begin{IEEEeqnarray}{rCl}
	\left(\tilde{\slashed{\nabla}}-\tilde{m}\right) \tilde{\psi} &=&\Omega^{-\frac{D+1}{2}}\left(\slashed{\nabla}-m\right)\psi=0  . \label{eq:realtionslashedtildepsislashedpsi}
\end{IEEEeqnarray} 
Therefore, we can consider the equivalent Dirac equation in the following conformally transformed background (with $\Omega=1/r$)
\begin{IEEEeqnarray}{rCl}
	\dd\tilde{s}^2 &=&\frac{1}{r^2}\dd s^2=-\frac{f}{r^2}\dd v^2 +\frac{2}{r^2}\dd v\dd r+\dd\Sigma_{D-2}^2  . \label{eq:conformalmetricspherical}
\end{IEEEeqnarray} 
Correspondingly, the tetrads can be chosen as follows 
\begin{IEEEeqnarray}{rCl}
	\tilde{e}_{v}^{~\mu} &=& r \left[\left(\frac{\partial}{\partial v}\right)^\mu+\frac{f(r)-1}{2}\left(\frac{\partial}{\partial r}\right)^{\mu}\right],\IEEEyesnumber \IEEEyessubnumber \label{eq:tetradDdimetmu}\\
	\tilde{e}_{r}^{~\mu} &=& r \left[\left(\frac{\partial}{\partial v}\right)^\mu+\frac{f(r)+1}{2}\left(\frac{\partial}{\partial r}\right)^{\mu}\right] , \IEEEyessubnumber\label{eq:tetradDdimermu} \\ 
	\tilde{e}_1^{~\mu} &=&\left(\frac{\partial}{\partial \phi^1}\right)^{\mu},~ \tilde{e}_i^{~\mu} =\frac{1}{\prod_{j=1}^{i-1}\sin\phi^j}\left(\frac{\partial}{\partial\phi^i}\right)^{\mu}  \IEEEyessubnumber\label{eq:tetradDdime12mu}
\end{IEEEeqnarray} 
with $i=2,~\cdots,~D-2$. 
The tetrads~\eqref{eq:tetradDdimetmu}-\eqref{eq:tetradDdime12mu} are finite on the outer horizon, as a result, the spinor field $\tilde{\psi}$ would be regular at $r=r_+$ provides its components are non-singular there.

\subsection{The gamma matrices and variable separation}
\label{sec:gammamatrices}
In $D$-dimensional spacetime, the gamma matrices can be given explicitly depending on whether the dimension $D$ is even or odd~\cite{Lopez-Ortega:2009flo,Liu:2019rbq}. For even $D$, 
\begin{IEEEeqnarray}{rCl}
	\gamma^v &=& \ii \sigma_1\otimes I_{2^{(D-2)/2}} , ~\gamma^r=\sigma_2\otimes I_{2^{(D-2)/2}}, ~\gamma^i=\sigma_2\otimes \hat{\gamma}^i  \label{eq:gammamatricesevenD}
\end{IEEEeqnarray} 
with $i=1,~\cdots,~D-2$, where 
\begin{IEEEeqnarray}{rCl}
	\hat{\gamma}^{2n-1} &=&\sigma_2^{\otimes (n-1)}\otimes \sigma_1\otimes I_{2^{(D-2)/2-n}} , ~\hat{\gamma}^{2n}=\sigma_2^{\otimes (n-1)}\otimes \sigma_2\otimes I_{2^{(D-2)/2-n}} 
\end{IEEEeqnarray} 
with $n=1,~\cdots,~(D-2)/2$. Similarly, for odd $D$, the gamma matrices are chosen as 
\begin{IEEEeqnarray}{rCl}
	\gamma^v &=& \ii \sigma_1\otimes I_{2^{(D-3)/2}} , ~\gamma^r=\sigma_2\otimes I_{2^{(D-3)/2}}, ~\gamma^i=\sigma_2\otimes \hat{\gamma}^i 
\end{IEEEeqnarray} 
with $i=1,~\cdots,~D-2$, where 
\begin{IEEEeqnarray}{rCl}
	\hat{\gamma}^{2n-1} &=&\sigma_2^{\otimes (n-1)}\otimes \sigma_1\otimes I_{2^{(D-3)/2-n}} , ~\hat{\gamma}^{2n}=\sigma_2^{\otimes (n-1)}\otimes \sigma_2\otimes I_{2^{(D-3)/2-n}} ,~\hat{\gamma}^{D-2}=\sigma_3^{\otimes (D-3)/2} \nonumber \\
\end{IEEEeqnarray} 
with $n=1,~\cdots,~(D-3)/2$. It is noteworthy that $\{\hat{\gamma}^i\}$ denote the gamma matrices for a $(D-2)$-dimensional space with the signature $(+\cdots +)$.

With the tetrad choice specified in Eqs.~\eqref{eq:tetradDdimetmu}-\eqref{eq:tetradDdime12mu}, the angular part of the Dirac operator $\tilde{\slashed{\nabla}}$ can be separated out as follows~\cite{Lopez-Ortega:2009flo,Liu:2019rbq}
\begin{IEEEeqnarray}{rCl}
	\tilde{\slashed{\nabla}} &=& \left(\ii\sigma_1\tilde{e}_t^{~\mu}\nabla^{(2)}_{\mu}\right)\otimes I +\sigma_3 \otimes \left(\hat{\gamma}^i\tilde{e}_i^{~\mu}\nabla_{\mu}^{(D-2)}\right)=\slashed{\nabla}_2\otimes I +\sigma_3\otimes \slashed{\nabla}_{\Sigma} , 
\end{IEEEeqnarray} 
where $\slashed{\nabla}_2$ is the Dirac operator associated with the effective two-dimensional spacetime 
\begin{IEEEeqnarray}{rCl}
	\dd s_2^2 &=& -\frac{f}{r^2}\dd v^2 +\frac{2}{r^2} \dd v\dd r, \label{eq:metric2dvr}
\end{IEEEeqnarray} 
and $\slashed{\nabla}_{\Sigma}$ is the Dirac operator on the $(D-2)$-dimensional unit sphere $\Sigma_{D-2}$.
This decomposition motivates the separation ansatz as follows
\begin{IEEEeqnarray}{rCl}
	\tilde{\psi}(v,r,\phi^i) &=& \varphi(v,r)\otimes \chi(\phi^i) , 
\end{IEEEeqnarray} 
where $\varphi(t,r)$ is a two-component spinor field in the $2$-dimensional spacetime with metric~\eqref{eq:metric2dvr}, and $\chi(\phi^i)$ is the eigen-spinor field of the Dirac operator on the unit sphere $\Sigma_{D-2}$ satisfying
\begin{IEEEeqnarray}{rCl}
	\slashed{\nabla}_{\Sigma}\chi &=&  -\ii \lambda \chi.
\end{IEEEeqnarray} 
The eigenvalue $\lambda$ is given by~\cite{Camporesi:1995fb,Liu:2019rbq,Pang:2025myy}
\begin{IEEEeqnarray}{rCl}
	\lambda&=&  l+\frac{D-3}{2} , \label{eq:lambdaDrelation}
\end{IEEEeqnarray} 
with the total angular momentum $l=\frac{1}{2},~\frac{3}{2},~\cdots$\footnote{The sign of $\lambda$ can be fixed by choosing a spinor basis for $\chi$, as illustrated in the appendix~\ref{sec:lambdasign4} for the $4$-dimensional case.}. 
Consequently, the Dirac equation reduces to
\begin{IEEEeqnarray}{rCl}
	(\slashed{\nabla}_2-\tilde{m})\varphi &=& (\slashed{\nabla}_2-m r)\varphi= \ii\lambda\sigma_3\varphi , \label{eq:radialslashednabla2varphi}
\end{IEEEeqnarray} 
where we have used the conformal transformation~\eqref{eq:conformaltransformation} of the spinor mass, $\tilde{m}=mr$.

\section{The radial equation and the static fermionic \acrshort{tln}s}
\label{sec:Lovenumber}
In the effective two-dimensional $(v,r)$ spacetime~\eqref{eq:metric2dvr}, the gamma matrices can be given as follows (setting $D=2$ in Eq.~\eqref{eq:gammamatricesevenD})
\begin{IEEEeqnarray}{rCl}
	\gamma^v&=&  \ii\sigma_1,~\gamma^r=\sigma_2,
\end{IEEEeqnarray} 
from which we can get the spin connection 
\begin{IEEEeqnarray}{rCl}
	\Gamma_v &=& \frac{2f(r)-rf'(r)}{4r}\sigma_3 ,~\Gamma_r=-\frac{r}{2}\sigma_3 . \label{eq:spinconnectionvr}
\end{IEEEeqnarray} 
Substituting the following ansatz 
\begin{IEEEeqnarray}{rCl}
	\varphi&=& (\varphi_1,\varphi_2)=\left(\sqrt{\frac{r}{2f(r)}}R_-(r)\ee^{-\ii\omega v}, \sqrt{\frac{r}{2}}R_+(r)\ee^{-\ii\omega v} \right)  \label{eq:phiRmRpansatz}
\end{IEEEeqnarray} 
and spin connection~\eqref{eq:spinconnectionvr} into Eq.~\eqref{eq:radialslashednabla2varphi}, we get the radial equation as follows
\begin{IEEEeqnarray}{rCl}
	\left[-2\ii \omega +f(r)\frac{\dd}{\dd r}\right]R_-(r) &=& -\sqrt{f(r)}\left(\frac{\lambda}{r}+\ii m \right) R_+(r), \IEEEyesnumber \IEEEyessubnumber \label{eq:Rmpr}\\ 
	f(r) \frac{\dd}{\dd r}R_+(r)&=& - \sqrt{f(r)}\left(\frac{\lambda}{r}-\ii m \right)R_-(r) . \IEEEyessubnumber \label{eq:Rppr}
\end{IEEEeqnarray} 
Note that both the blackening factor $f(r)$ (Eq.~\eqref{eq:blackeningfactor}) and the eigenvalue $\lambda$ (Eq.~\eqref{eq:lambdaDrelation}) are $D$-dependent. Setting $m=0$ and $D=4$ we recover the radial equation in $4$-dimensional \acrshort{rn} spacetime~\cite{Pang:2025myy}. 

To solve the radial equations~\eqref{eq:Rmpr} and~\eqref{eq:Rppr}, we eliminate $R_+$ to obtain a second-order equation for $R_-$.
By introducing the following dimensionless variable
\begin{IEEEeqnarray}{rCl}
	z &=& \frac{r^{D-3}-r_+^{D-3}}{r_+^{D-3}-r_-^{D-3}} , \label{eq:defz}
\end{IEEEeqnarray} 
the second-order equation for $R_-(z)$ under massless and static perturbations ($m = \omega = 0$) becomes
\begin{IEEEeqnarray}{rCl}
	2z(1+z)R_-''(z)+(1+2z)R_-'(z) -2\tilde{\lambda}^2 R_-(z) &=&0  , \label{eq:Rmppzlmbdat}
\end{IEEEeqnarray} 
where we have introduced a new parameter $\tilde{\lambda}$
\begin{IEEEeqnarray}{rCl}
	\tilde{\lambda} &=& \frac{\lambda}{D-3}=\frac{l}{D-3}+\frac{1}{2}  , 
\end{IEEEeqnarray} 
and the prime denotes derivatives respect to the dimensionless variable $z$.
Eq.~\eqref{eq:Rmppzlmbdat} is identical to Eq.~(2.14) in our previous paper~\cite{Pang:2025myy} with $\lambda$ replaced by $\tilde{\lambda}$, whose general solution can be given as follows~\cite{Chakraborty:2025zyb,Pang:2025myy} 
\begin{IEEEeqnarray}{rCl}
	R_-(z) &=& c_1\sqrt{1+z}{}_2F_1 \left(\frac{1}{2}-\tilde{\lambda},\frac{1}{2}+\tilde{\lambda},\frac{1}{2},-z\right)+c_2\sqrt{z} {}_2F_1 \left(\frac{1}{2}-\tilde{\lambda},\frac{1}{2}+\tilde{\lambda},\frac{3}{2},-z\right) , \nonumber \\
\end{IEEEeqnarray} 
with ${}_2F_1$ the hypergeometric function~\cite{Gradshteyn:1996table}.

At the outer horizon, $z = 0$ (from Eq.~\eqref{eq:defz}), the blackening factor $f$ exhibits the following asymptotic behavior
\begin{IEEEeqnarray}{rCl}
	f(z) &\to& \frac{(r_+^{D-3}-r_-^{D-3})^2z}{r_+^{2(D-3)}}, \label{eq:fz0exphigherd}
\end{IEEEeqnarray} 
Therefore, when $c_1\neq 0$, the quantity $R_-/\sqrt{f}$ diverges at the horizon $z=0$ as follows
\begin{IEEEeqnarray*}{rCl}
	\frac{R_-(z)}{\sqrt{f(z)}} &\to& \frac{r_+^{D-3}}{\left(r_+^{D-3}-r_-^{D-3}\right)\sqrt{z}}c_1 +\frac{r_+^{D-3}}{r_+^{D-3}-r_-^{D-3}}c_2 . 
\end{IEEEeqnarray*} 
Regularity of the spinor field at the horizon then requires $c_1 = 0$. Since the static \acrshort{tln}s are independent of the overall factor $c_2$~\cite{Bhatt:2023zsy,Chakraborty:2025zyb}, we set $c_2 = 1$, yielding the regular solution~\cite{Pang:2025myy}
\begin{IEEEeqnarray}{rCl}
	R_-(z) &=& \sqrt{z} {}_2F_1 \left(\frac{1}{2}-\tilde{\lambda},\frac{1}{2}+\tilde{\lambda},\frac{3}{2},-z\right) , 
\end{IEEEeqnarray} 
which has the following asymptotic behavior when $z\rightarrow\infty$
\begin{align}
	R_-  &\to\frac{2^{2\tilde{\lambda}-2}\Gamma\left(-\tilde{\lambda}\right)}{\Gamma(\tilde{\lambda}+1)} z^{\tilde{\lambda}}\left[(1+\cdots)+\frac{\Gamma\left(-\tilde{\lambda}\right)\Gamma\left(1+\tilde{\lambda}\right)}{4^{2\tilde{\lambda}}\Gamma\left(1-\tilde{\lambda}\right)\Gamma\left(\tilde{\lambda}\right)}z^{-2\tilde{\lambda}}(1+\cdots)\right],
    \nonumber \\ 
		 &\to -\frac{4^{\tilde{\lambda}-1}}{\tilde{\lambda}}  \frac{r^{l+\frac{D-3}{2}}}{(r_+^{D-3}-r_-^{D-3})^{\tilde{\lambda}}} \left\{(1+\cdots)-\frac{1}{4^{2\tilde{\lambda}}}\left[1-\left(\frac{r_-}{r_+}\right)^{D-3}\right]^{2\tilde{\lambda}}\left(\frac{r_+}{r}\right)^{2l+(D-3)}(1+\cdots)\right\}.\label{eq:Rmsolrnexphigherd}
\end{align}
The static \acrshort{tln} $\mathcal{F}_{-\frac{1}{2}lm}$ is defined as the coefficient of the $\left(\frac{r_+}{r}\right)^{2l+(D-3)}$ term in the curly bracket~\cite{Hui:2020xxx}.
Similarly, the static \acrshort{tln} $\mathcal{F}_{\frac{1}{2}lm}$ for spin‑up perturbations is extracted from the large‑$z$ expansion of $R_+$. The results read
\begin{IEEEeqnarray}{rCl}
	\mathcal{F}_{\pm\frac{1}{2}lm} &=& \pm\frac{1}{4^{2\tilde{\lambda}}}\Biggl[1-\Bigl(\frac{r_-}{r_+}\Bigr)^{D-3}\Biggr]^{2\tilde{\lambda}}
	= \mathcal{F}_{\pm \frac{1}{2}lm}^{\text{Sch}}\left[\frac{2\sqrt{1-q^2}}{1+\sqrt{1-q^2}}\right]^{\frac{2l}{D-3}+1},  \label{eq:Fpm1p2D}
\end{IEEEeqnarray} 
where $q = Q/M$ is the dimensionless charge parameter, and
\begin{IEEEeqnarray}{rCl}
	\mathcal{F}_{0 \frac{1}{2}lm}^{\text{Sch}} &=& \pm 4^{-\frac{2l}{D-3}-1}  . \label{eq:Fpm1p2DSch}
\end{IEEEeqnarray}
denotes the fermionic TLNs for Schwarzschild black holes ($q=0$). For completeness, the static response of a neutral scalar in $D$ dimensions is given by~\cite{Ma:2024few,Rai:2024lho}
\begin{IEEEeqnarray}{rCl}
    \mathcal{F}_{\pm \frac{1}{2}lm}^{\text{Sch}} &=& \frac{2\tilde{l}+1}{2\pi}\frac{\Gamma(\tilde{l}+1)}{\Gamma(2\tilde{l}+2)^2}\tan(\pi\tilde{l}),\quad \tilde{l}=\frac{l}{D-3}.
\end{IEEEeqnarray}
For bosonic perturbations, the total angular momentum $l$ is an integer, and the static \acrshort{tln}s vanish identically in $4$-dimension spacetimes; while for $D>4$, they are generically non‑zero except when $\tilde{l}=l/(D-3)$ is an integer. This behavior contrasts sharply with the fermionic case: static fermionic \acrshort{tln}s remain non‑zero for all dimensions $D \geq 4$ and total angular momentum $l \geq \frac{1}{2}$\footnote{For fermionic perturbations, the total angular momentum $l$ is a half-integer.}.

For a fixed spacetime dimension $D \geq 4$, the magnitude of the static fermionic \acrshort{tln}s, $\left|\mathcal{F}_{\pm \frac{1}{2}lm}\right|$, decreases monotonically as $q$ or $l$ increases, vanishing only in the extremal limit $q \to 1$\footnote{Appendix~\ref{sec:extremaltln} shows that the static fermionic \acrshort{tln}s are indeed vanish for extremal \acrshort{rn} black holes.}. 
Furthermore, as $D$ increases, the sensitivity of the static \acrshort{tln}s to $l$ diminishes. In the limit $D \to \infty$, the static fermionic \acrshort{tln}s become independent of the total angular momentum $l$.

\section{Discussion}
\label{sec:discussion}

In this paper, we have extended the study of static fermionic \acrshort{tln}s to higher-dimensional \acrshort{rn} black holes. By employing the ingoing Eddington coordinate and appropriate tetrads that ensure regularity at the horizon, we separated the angular part of the Dirac operator through a conformal transformation and obtained the reduced equation in an effective two-dimensional spacetime. The resulting radial equation for massless and static fermionic perturbations in $D$ dimensions maintains a form similar to the four-dimensional case, but with a crucial dimension-dependent parameter $\tilde{\lambda} = \frac{l}{D-3} + \frac{1}{2}$. 

Correspondingly, the static fermionic \acrshort{tln}s are also dimension-dependent, and are non-vanishing $l \geq 1/2$ except for the extremal case. This is in sharp contrast to static bosonic \acrshort{tln}s, which vanish identically for $D=4$ and for integer $\frac{l}{D-3}$ when $D>4$.
We also note that as the spacetime dimension $D$ increases, the sensitivity of static \acrshort{tln}s to the total angular momentum $l$ decreases, with the static \acrshort{tln}s becoming independent of $l$ in the limit $D \to \infty$.
These results demonstrate that fermionic tidal responses exhibit a rich structure that depends critically on both the spacetime dimension and the black hole charge, and opens new avenues for understanding fermionic tidal deformability in higher-dimensional spacetimes. 

We conclude by outlining several promising directions for future research. First, our analysis does not extend to $3$-dimensional spacetimes~\cite{Martinez:1999qi,Bhatt:2024mvr}; investigating fermionic tidal responses of proper $3$-dimensional black holes would also be interesting. Second, it would be valuable to extend our work to rotating black holes in higher dimensions, examine the effects of fermion mass on the static \acrshort{tln}s, and explore implications for gravitational wave astronomy in theories with extra dimensions. Last but not least, the connection between static fermionic \acrshort{tln}s and black hole quantum properties merits further investigation to determine whether fermionic responses can probe quantum gravity effects. These potential generalizations will be reported elsewhere in the future.

\acknowledgments

This work is partially supported by the National Key Research and Development Program of China with Grant No. 2021YFC2203001 as well as the National Natural Science Foundation of China (NSFC) with Grant Nos. 12035016, 12275350, 12375048, 12375058, 12361141825, 12447182, 12575047, and 12505082. 
XP is also supported by the Doctoral Initiation Grant 24KE051 and  Basic Research Grant 25kx010 from China West Normal University. QJ is supported by the Key Joint Program of Science and Education of Sichuan Province with Grant No. 25LHJJ0097.

\appendix 
\section{The Dirac equation in the conformal transformed background}
\label{sec:diracequationconformal}
In this appendix we will prove the transformation property~\eqref{eq:realtionslashedtildepsislashedpsi} of the Dirac equation under the conformal transformations~\eqref{eq:conformaltransformation}.

\paragraph{The Ricci rotation coefficients.} In the conformal transformed spacetime, the Ricci rotation coefficients read
\begin{IEEEeqnarray}{rCl}
	\tilde{\Gamma}_{abc} &=& \left(\tilde{e}_{a\mu}\right)_{;\nu}\tilde{e}_b^{~\mu}\tilde{e}_c^{~\nu}=\left(\partial_{\nu} \tilde{e}_{a\mu} - \tilde{\Gamma}^{\lambda}_{\nu\mu}\tilde{e}_{a\lambda} \right)\tilde{e}_b^{~\mu}\tilde{e}_c^{~\nu},
\end{IEEEeqnarray} 
with $\tilde{\Gamma}^{\lambda}_{\nu\mu}$ are the Christoffel symbols calculated using the conformal transformed metric $\tilde{g}_{\mu\nu}$
\begin{IEEEeqnarray*}{rCl}
	\tilde{\Gamma}^{\lambda}_{\nu\mu} &=& \frac{1}{2}\tilde{g}^{\lambda\sigma}\left(\frac{\partial \tilde{g}_{\sigma\mu}}{\partial x^{\nu}}+\frac{\partial \tilde{g}_{\sigma\nu}}{\partial x^{\mu}}-\frac{\partial \tilde{g}_{\nu\mu}}{\partial x^{\sigma}}\right) = \Gamma^{\lambda}_{\nu\mu} + \frac{1}{\Omega}\left[(\partial_{\nu}\Omega )\delta^{\lambda}_{\mu} +(\partial_{\mu}\Omega )\delta^{\lambda}_{\nu}-(\partial_{\sigma}\Omega) g^{\lambda\sigma}g_{\nu\mu}\right] ,  \\
\end{IEEEeqnarray*} 
where we have used 
\begin{IEEEeqnarray*}{rCl}
	\frac{\partial \tilde{g}_{\sigma\mu}}{\partial x^{\nu}} &=& \frac{\partial}{\partial x^{\nu}} \left(\Omega^2 g_{\sigma\mu}\right) =2\Omega (\partial_{\nu}\Omega) g_{\sigma\mu} +\Omega^2 \partial_{\nu} g_{\sigma\mu} . 
\end{IEEEeqnarray*} 
Then one has 
\begin{IEEEeqnarray*}{rCl}
	\left(\tilde{e}_{a\mu}\right)_{;\nu}&=&\left(\partial_{\nu} \tilde{e}_{a\mu} - \tilde{\Gamma}^{\lambda}_{\nu\mu}\tilde{e}_{a\lambda} \right)\tilde{e}_b^{~\mu}\tilde{e}_c^{~\nu} , \\
	&=&\partial_\nu \left(\Omega e_{a\mu}\right)-\tilde{\Gamma}^{\lambda}_{\nu\mu} \tilde{e}_{a\lambda} , \\ 
								  &=& \Omega \left(\nabla_{\nu}e_{a\mu}\right) +(\partial_{\sigma}\Omega )e_a^{~\sigma}g_{\nu\mu} -(\partial_{\mu}\Omega )e_{a\nu} , 
\end{IEEEeqnarray*} 
where we have used the fact that $e_{a\lambda}=\Omega^{-1}\tilde{e}_{a\lambda}$.
Submitting the result into $\tilde{\Gamma}_{abc}$, we obtain 
\begin{IEEEeqnarray}{rCl}
	\tilde{\Gamma}_{abc} &=& \frac{1}{\Omega}\Gamma_{abc} +\frac{1}{\Omega^2}\left[\vec{e}_a\left(\Omega\right)\eta_{bc}-\vec{e}_b\left(\Omega\right)\eta_{ac}\right] , \label{eq:tildeGammaldldldGammaldldld} 
\end{IEEEeqnarray} 
with $\vec{e}_a(\Omega)=e_a^{~\mu}(\partial_{\mu}\Omega)$.
\paragraph{The Dirac equation.} To prove the transformation property~\eqref{eq:realtionslashedtildepsislashedpsi} of the Dirac equation, we first note that 
\begin{IEEEeqnarray}{rCl}
	\Gamma_c &=&-\frac{\ii}{2}\Gamma_{abc}\Sigma^{ab}=\frac{1}{4}\Gamma_{abc}\gamma^a\gamma^b  , 
\end{IEEEeqnarray} 
then
\begin{IEEEeqnarray*}{rCl}
	\tilde{\gamma}^{\mu}\tilde{\Gamma}_{\mu}=\tilde{\gamma}^c\tilde{\Gamma}_c &=&\frac{\tilde{\gamma}^c}{4}\tilde{\Gamma}_{abc}\tilde{\gamma}^a\tilde{\gamma}^b  = \frac{\gamma^c\Gamma_c}{\Omega}-\frac{D-1}{2\Omega^2}\vec{e}_a(\Omega)\gamma^a ,  
\end{IEEEeqnarray*} 
where we have used that 
\begin{IEEEeqnarray*}{rCl}
	\gamma^c \vec{e}_a(\Omega)\eta_{bc}\gamma^a\gamma^b-\gamma^c \vec{e}_b(\Omega)\eta_{ac}\gamma^a\gamma^b &=& \vec{e}_a(\Omega)\left(\{\gamma^c,\gamma^a\}-\gamma^a\gamma^c\right)\gamma^b\eta_{bc}-\vec{e}_b(\Omega)\eta_{ac}\gamma^c\gamma^a\gamma^b  , \\
																											   &=& \vec{e}_a(\Omega)(2\eta^{ca})\gamma^b\eta_{bc}-\vec{e}_a(\Omega)\eta_{bc}\gamma^a\gamma^c\gamma^b-\vec{e}_b(\Omega)\eta_{ac}\gamma^a\gamma^c\gamma^b ,  \\ 
																											   &=& 2(1-D)\vec{e}_a(\Omega)\gamma^a .  
\end{IEEEeqnarray*} 
In the last line, the result $\eta_{bc}\gamma^b\gamma^c=\frac{1}{2}\eta_{bc}\{\gamma^b,\gamma^c\}=\eta_{bc}\eta^{bc}=D$ is used.
Setting $\tilde{\psi}=\Omega^s\psi$, and note that fact that $\tilde{\vec{e}}_c(h)=\tilde{e}_c^{~\mu}\partial_{\mu}h=\frac{1}{\Omega}e_c^{~\mu}\partial_{\mu}h=\frac{\vec{e}_c(h)}{\Omega}$ for arbitrary function $h$, and 
\begin{IEEEeqnarray*}{rCl}
	\vec{e}_c\tilde{\psi} &=& \vec{e}_c \left(\Omega^s\psi\right) =s \Omega^{s-1}\vec{e}_c\left(\Omega\right)\psi +\Omega^s \left(\vec{e}_c\psi\right), 
\end{IEEEeqnarray*} 
we obtain 
\begin{IEEEeqnarray*}{rCl}
	\tilde{\slashed{\nabla}}\tilde{\psi} &=& \tilde{\gamma^c}\left(\tilde{\vec{e}}_c-\tilde{\Gamma}_c\right)\tilde{\psi}, \\ 
										 &=& \gamma^c \frac{\vec{e}_c}{\Omega}\tilde{\psi}-\tilde{\gamma}^c\tilde{\Gamma}_c\tilde{\psi} , \\   
										 &=& \frac{\gamma^c}{\Omega}\left[s\Omega^{s-1}\vec{e}_c(\Omega)\psi +\Omega^s \left(\vec{e}_c\psi\right)\right] - \frac{\gamma^c\Gamma_c}{\Omega}\Omega^s\psi +\frac{D-1}{2\Omega^2}\vec{e}_a(\Omega)\gamma^a\Omega^s\psi , \\  
										 &=& \Omega^{s-1}\slashed{\nabla}\psi + \Omega^{s-2}\vec{e}_a(\Omega)\gamma^a\psi \left(s+\frac{D-1}{2}\right) .  
\end{IEEEeqnarray*} 
Therefore, letting $\tilde{\slashed{\nabla}}\tilde{\psi}$ proportional to $\slashed{\nabla}\psi$ we would require 
\begin{IEEEeqnarray*}{rCl}
	s &=& -\frac{D-1}{2} , 
\end{IEEEeqnarray*} 
and then $\tilde{\psi}=\Omega^{-\frac{D-1}{2}}\psi$ results in 
\begin{IEEEeqnarray*}{rCl}
	\left(\tilde{\slashed{\nabla}}-\tilde{m}\right)\tilde{\psi} &=& \Omega^{-\frac{D+1}{2}} \left(\slashed{\nabla}-m\right)\psi  
\end{IEEEeqnarray*} 
if we set $\tilde{m}=\Omega^{-1}m$, and the relation~\eqref{eq:realtionslashedtildepsislashedpsi} is proved.

\section{The angular Dirac operator in $4$-dimension}
\label{sec:lambdasign4}
In this appendix we will illustrate how to fix the sign of the eigenvalue $\lambda$. Setting $D=4$ one can get the angular Dirac operator on the $2$-sphere reads
\begin{IEEEeqnarray}{rCl}
	\slashed{\nabla}_{\Sigma} &=& -\left(\begin{array}{cc}
		0 & \bar{\eth} \\
\eth & 0 \\ \end{array}\right) , 
\end{IEEEeqnarray} 
where $\eth$ and $\bar{\eth}$ are raising and lowering operator on spin-weighted spherical harmonics ${}_s Y_{lm}$~\cite{Goldberg:1966uu,Penrose:1985bww}, respectively. When acting on ${}_{\pm \frac{1}{2}}Y_{lm}$, we have~\cite{Pang:2025myy}
\begin{IEEEeqnarray}{rCl}
	\bar{\eth}{}_{\frac{1}{2}}Y_{lm} ( \phi_1,\phi_2)&=& -\left(\partial_{\phi_1}-\frac{\ii}{\sin\phi_1}\partial_{\phi_2}+\frac{1}{2}\cot{\phi_1}\right) {}_{\frac{1}{2}}Y_{lm}(\phi_1,\phi_2) = -\lambda {}_{-\frac{1}{2}}Y_{lm}(\phi_1,\phi_2), \nonumber \\ \\
	\eth{}_{-\frac{1}{2}}Y_{lm} ( \phi_1,\phi_2)&=& -\left(\partial_{\phi_1}+\frac{\ii}{\sin\phi_1}\partial_{\phi_2}+\frac{1}{2}\cot{\phi_1}\right) {}_{-\frac{1}{2}}Y_{lm}(\phi_1,\phi_2) = \lambda {}_{-\frac{1}{2}}Y_{lm}(\phi_1,\phi_2), \nonumber \\
\end{IEEEeqnarray} 
for $\lambda=l+\frac{1}{2}$.
Letting 
\begin{IEEEeqnarray}{rCl}
	\chi &=& \left(\begin{array}{c}
		\ii S_-(\theta,\varphi) \\ S_+(\theta,\varphi)
	\end{array}\right) =\left(\begin{array}{c}
	\ii {}_{-\frac{1}{2}}Y_{lm} \\{}_{\frac{1}{2}}Y_{lm}
	\end{array}\right), 
\end{IEEEeqnarray} 
one can show that 
\begin{IEEEeqnarray}{rCl}
	\slashed{\nabla}_{\Sigma}\chi &=& \left(\begin{array}{c}
			 -\bar{\eth}S_+ \\ \ii\eth S_- 
	\end{array}\right) =\left(\begin{array}{c}
		\lambda S_- \\ \ii \lambda S_+
	\end{array}\right) = -\ii \lambda \left(\begin{array}{c}
		\ii S_- \\ S_+
	\end{array}\right), 
\end{IEEEeqnarray} 
therefore the eigenvalue of $\slashed{\nabla}_{\Sigma}$ is $-\ii \lambda$ with $\lambda=l+\frac{1}{2}$, exactly $l+\frac{D-3}{2}$ for $D=4$.

\section{The static fermionic response of the extremal RN black hole}
\label{sec:extremaltln}
For the extremal \acrshort{rn} black holes with $Q=M$, the blackening factor $f(r)$ simplifies to a total square 
\begin{IEEEeqnarray}{rCl}
	f(r) &=& \left(\frac{r^{D-3}-M}{r^{D-3}}\right)^2 =\left(\frac{\tilde{z}}{1+\tilde{z}}\right)^2, 
\end{IEEEeqnarray} 
where we have introduced a new dimensionless variable 
\begin{IEEEeqnarray}{rCl}
	\tilde{z} &=& \frac{r^{D-3}-M}{M} . 
\end{IEEEeqnarray} 
The second-order equation of $R_-$ for the massless and static perturbations becomes 
\begin{IEEEeqnarray}{rCl}
	\tilde{z}^2 \frac{\dd^2}{\dd \tilde{z}^2}R_-(\tilde{z})+\tilde{z} \frac{\dd}{\dd \tilde{z}}R_-(\tilde{z})-\tilde{\lambda}^2 R_-(\tilde{z}) &=& 0 , 
\end{IEEEeqnarray} 
whose solutions that are regular at the horizon $\tilde{z}=0$ can be written as follows 
\begin{IEEEeqnarray}{rCl}
	R_-(\tilde{z}) &=& \tilde{z}^{\tilde{\lambda}}.
\end{IEEEeqnarray} 
At infinity, the solution has the following asymptotic behaviour 
\begin{IEEEeqnarray}{rCl}
	R_-(r) &\to&\frac{1}{M^{\frac{l}{D-3}+\frac{1}{2}}}r^{l+\frac{D-3}{2}} , 
\end{IEEEeqnarray} 
where the absence of the $\left(\frac{r_+}{r}\right)^{2l+(D-3)}$ term indicates $\mathcal{F}_{-\frac{1}{2}lm}=0$ for extremal \acrshort{rn} black holes. The same behaviour applies to $\mathcal{F}_{\frac{1}{2}lm}$.

\printglossary

\input{./main.bbl}

\end{document}

%% file: main.bbl
\providecommand{\href}[2]{#2}\begingroup\raggedright\endgroup

%% file: main.bbl
\begin{thebibliography}{10}

\bibitem{Love:1909yed}
A.E.H.~Love, \emph{The yielding of the earth to disturbing forces}, \href{https://doi.org/10.1098/rspa.1909.0008}{\emph{Proc. R. Soc. Lond. A} {\bfseries 82} (1909) 73}.

\bibitem{Binnington:2009bb}
T.~Binnington and E.~Poisson, \emph{Relativistic theory of tidal {{Love}} numbers}, \href{https://doi.org/10.1103/PhysRevD.80.084018}{\emph{Phys. Rev. D} {\bfseries 80} (2009) 084018} [\href{https://arxiv.org/abs/0906.1366}{{\ttfamily 0906.1366}}].

\bibitem{Bhatt:2023zsy}
R.P.~Bhatt, S.~Chakraborty and S.~Bose, \emph{Addressing issues in defining the {{Love}} numbers for black holes}, \href{https://doi.org/10.1103/PhysRevD.108.084013}{\emph{Phys. Rev. D} {\bfseries 108} (2023) 084013} [\href{https://arxiv.org/abs/2306.13627}{{\ttfamily 2306.13627}}].

\bibitem{Rodriguez:2026iot}
M.J.~Rodr{\'i}guez, L.~Santoni and A.R.~Solomon, \emph{Love numbers of black holes and compact objects},  Apr., 2026.
\newblock arXiv.2604.08653.

\bibitem{Chakraborty:2026qru}
S.~Chakraborty and P.~Pani, \emph{{Tidal Response of Compact Objects}},  \href{https://arxiv.org/abs/2604.08679}{{\ttfamily 2604.08679}}.

\bibitem{Chakravarti:2018vlt}
K.~Chakravarti, S.~Chakraborty, S.~Bose and S.~SenGupta, \emph{Tidal {{Love Numbers}} of {{Black Holes}} and {{Neutron Stars}} in the {{Presence}} of {{Higher Dimensions}}: {{Implications}} of {{ GW170817}}}, \href{https://doi.org/10.1103/PhysRevD.99.024036}{\emph{Phys. Rev. D} {\bfseries 99} (2019) 024036} [\href{https://arxiv.org/abs/1811.11364}{{\ttfamily 1811.11364}}].

\bibitem{LeTiec:2020spy}
A.L.~Tiec and M.~Casals, \emph{Spinning {{Black Holes Fall}} in {{Love}}}, \href{https://doi.org/10.1103/PhysRevLett.126.131102}{\emph{Phys. Rev. Lett.} {\bfseries 126} (2021) 131102} [\href{https://arxiv.org/abs/2007.00214}{{\ttfamily 2007.00214}}].

\bibitem{Flanagan:2007ix}
E.E.~Flanagan and T.~Hinderer, \emph{Constraining neutron star tidal {{Love}} numbers with gravitational wave detectors}, \href{https://doi.org/10.1103/PhysRevD.77.021502}{\emph{Phys. Rev. D} {\bfseries 77} (2008) 021502} [\href{https://arxiv.org/abs/0709.1915}{{\ttfamily 0709.1915}}].

\bibitem{Hinderer:2007mb}
T.~Hinderer, \emph{Tidal {{Love}} numbers of neutron stars}, \href{https://doi.org/10.1086/533487}{\emph{Astrophys. J.} {\bfseries 677} (2008) 1216} [\href{https://arxiv.org/abs/0711.2420}{{\ttfamily 0711.2420}}].

\bibitem{Damour:2009vw}
T.~Damour and A.~Nagar, \emph{Relativistic tidal properties of neutron stars}, \href{https://doi.org/10.1103/PhysRevD.80.084035}{\emph{Phys. Rev. D} {\bfseries 80} (2009) 084035} [\href{https://arxiv.org/abs/0906.0096}{{\ttfamily 0906.0096}}].

\bibitem{Chatziioannou:2020pqz}
K.~Chatziioannou, \emph{Neutron star tidal deformability and equation of state constraints}, \href{https://doi.org/10.1007/s10714-020-02754-3}{\emph{Gen. Rel. Grav.} {\bfseries 52} (2020) 109} [\href{https://arxiv.org/abs/2006.03168}{{\ttfamily 2006.03168}}].

\bibitem{Poisson:2014gka}
E.~Poisson, \emph{Tidal deformation of a slowly rotating black hole}, \href{https://doi.org/10.1103/PhysRevD.91.044004}{\emph{Phys. Rev. D} {\bfseries 91} (2015) 044004} [\href{https://arxiv.org/abs/1411.4711}{{\ttfamily 1411.4711}}].

\bibitem{Landry:2015zfa}
P.~Landry and E.~Poisson, \emph{Tidal deformation of a slowly rotating material body. {{External}} metric}, \href{https://doi.org/10.1103/PhysRevD.91.104018}{\emph{Phys. Rev. D} {\bfseries 91} (2015) 104018} [\href{https://arxiv.org/abs/1503.07366}{{\ttfamily 1503.07366}}].

\bibitem{Pani:2015hfa}
P.~Pani, L.~Gualtieri, A.~Maselli and V.~Ferrari, \emph{Tidal deformations of a spinning compact object}, \href{https://doi.org/10.1103/PhysRevD.92.024010}{\emph{Phys. Rev. D} {\bfseries 92} (2015) 024010} [\href{https://arxiv.org/abs/1503.07365}{{\ttfamily 1503.07365}}].

\bibitem{LeTiec:2020bos}
A.L.~Tiec, M.~Casals and E.~Franzin, \emph{Tidal {{Love Numbers}} of {{Kerr Black Holes}}}, \href{https://doi.org/10.1103/PhysRevD.103.084021}{\emph{Phys. Rev. D} {\bfseries 103} (2021) 084021} [\href{https://arxiv.org/abs/2010.15795}{{\ttfamily 2010.15795}}].

\bibitem{Chia:2020yla}
H.S.~Chia, \emph{Tidal {{Deformation}} and {{Dissipation}} of {{Rotating Black Holes }}}, \href{https://doi.org/10.1103/PhysRevD.104.024013}{\emph{Phys. Rev. D} {\bfseries 104} (2021) 024013} [\href{https://arxiv.org/abs/2010.07300}{{\ttfamily 2010.07300}}].

\bibitem{Hui:2021vcv}
L.~Hui, A.~Joyce, R.~Penco, L.~Santoni and A.R.~Solomon, \emph{Ladder {{Symmetries}} of {{Black Holes}}: {{Implications}} for {{ Love Numbers}} and {{No-Hair Theorems}}}, \href{https://doi.org/10.1088/1475-7516/2022/01/032}{\emph{JCAP} {\bfseries 01} (2022) 032} [\href{https://arxiv.org/abs/2105.01069}{{\ttfamily 2105.01069}}].

\bibitem{Hui:2022vbh}
L.~Hui, A.~Joyce, R.~Penco, L.~Santoni and A.R.~Solomon, \emph{{Near-zone symmetries of Kerr black holes}}, \href{https://doi.org/10.1007/JHEP09(2022)049}{\emph{JHEP} {\bfseries 09} (2022) 049} [\href{https://arxiv.org/abs/2203.08832}{{\ttfamily 2203.08832}}].

\bibitem{BenAchour:2022uqo}
J.~Ben~Achour, E.R.~Livine, S.~Mukohyama and J.-P.~Uzan, \emph{{Hidden symmetry of the static response of black holes: applications to Love numbers}}, \href{https://doi.org/10.1007/JHEP07(2022)112}{\emph{JHEP} {\bfseries 07} (2022) 112} [\href{https://arxiv.org/abs/2202.12828}{{\ttfamily 2202.12828}}].

\bibitem{Sharma:2024hlz}
C.~Sharma, R.~Ghosh and S.~Sarkar, \emph{Exploring {{Ladder Symmetry}} and {{Love Numbers}} for {{Static}} and {{Rotating Black Holes}}}, \href{https://doi.org/10.1103/PhysRevD.109.L041505}{\emph{Phys. Rev. D} {\bfseries 109} (2024) L041505} [\href{https://arxiv.org/abs/2401.00703}{{\ttfamily 2401.00703}}].

\bibitem{Rai:2024lho}
M.~Rai and L.~Santoni, \emph{Ladder {{Symmetries}} and {{Love Numbers}} of {{Reissner--Nordström Black Holes}}}, \href{https://doi.org/10.1007/JHEP07(2024)098}{\emph{JHEP} {\bfseries 07} (2024) 098} [\href{https://arxiv.org/abs/2404.06544}{{\ttfamily 2404.06544}}].

\bibitem{Combaluzier-Szteinsznaider:2024sgb}
O.~{Combaluzier-Szteinsznaider}, L.~Hui, L.~Santoni, A.R.~Solomon and S.S.C.~Wong, \emph{Symmetries of vanishing nonlinear {{Love}} numbers of {{ Schwarzschild}} black holes}, \href{https://doi.org/10.1007/JHEP03(2025)124}{\emph{Journal of High Energy Physics} {\bfseries 2025} (2025) 124} [\href{https://arxiv.org/abs/2410.10952}{{\ttfamily 2410.10952}}].

\bibitem{Berens:2025lsh}
R.~Berens, L.~Hui, D.~McLoughlin, A.R.~Solomon and J.~Staunton, \emph{Ladder {{Symmetries}} of {{Higher Dimensional Black Holes}}},  \href{https://arxiv.org/abs/2510.26748}{{\ttfamily 2510.26748}}.

\bibitem{DeLuca:2025zqr}
V.~De~Luca, B.~Khek, J.~Khoury and M.~Trodden, \emph{Hidden symmetries for tidal {{Love}} numbers: {{Generalities}} and applications to analog black holes}, \href{https://doi.org/10.1103/5bdk-xclx}{\emph{Phys. Rev. D} {\bfseries 113} (2026) 044006}.

\bibitem{Cvetic:2026wht}
M.~Cveti{\v c}, M.A.~Liao and M.M.~Stetsko, \emph{Tidal perturbations and {{Love}} symmetry for five-dimensional charged rotating black holes}, \href{https://doi.org/10.1103/xtcv-bsdn}{\emph{Phys. Rev. D} {\bfseries 113} (2026) 085008}.

\bibitem{Ghosh:2026vig}
R.~Ghosh, R.P.~Bhatt, S.~Chakraborty and S.~Bose, \emph{{Universal Ladder Structure Across Scales: From Quantum to Black Hole Physics}},  \href{https://arxiv.org/abs/2604.06249}{{\ttfamily 2604.06249}}.

\bibitem{Chakraborty:2023zed}
S.~Chakraborty, E.~Maggio, M.~Silvestrini and P.~Pani, \emph{Dynamical tidal {{Love}} numbers of {{Kerr-like}} compact objects}, \href{https://doi.org/10.1103/PhysRevD.110.084042}{\emph{Physical Review D} {\bfseries 110} (2024) 084042}.

\bibitem{Chakraborty:2025wvs}
S.~Chakraborty, V.D.~Luca, L.~Gualtieri and P.~Pani, \emph{Dynamical {{Love}} numbers of black holes: Theory and gravitational waveforms},  \href{https://arxiv.org/abs/2507.22994}{{\ttfamily 2507.22994}}.

\bibitem{Combaluzier--Szteinsznaider:2025eoc}
O.~Combaluzier-Szteinsznaider, D.~Glazer, A.~Joyce, M.J.~Rodriguez and L.~Santoni, \emph{{Dynamical tidal response of Schwarzschild Black Holes}}, \href{https://doi.org/10.1007/JHEP06(2026)032}{\emph{JHEP} {\bfseries 06} (2026) 032} [\href{https://arxiv.org/abs/2511.02372}{{\ttfamily 2511.02372}}].

\bibitem{Solon:2026ubm}
M.P.~Solon, \emph{{Universal Closed Form for Dynamical Love Numbers of Black Holes}},  \href{https://arxiv.org/abs/2606.19281}{{\ttfamily 2606.19281}}.

\bibitem{Saketh:2026trm}
M.V.S.~Saketh, S.~Ghosh and N.~Andersson, \emph{{Dynamical tidal response of neutron stars via scattering amplitudes}},  \href{https://arxiv.org/abs/2606.14405}{{\ttfamily 2606.14405}}.

\bibitem{Cardoso:2018ptl}
V.~Cardoso, M.~Kimura, A.~Maselli and L.~Senatore, \emph{Black {{Holes}} in an {{Effective Field Theory Extension}} of {{ General Relativity}}}, \href{https://doi.org/10.1103/PhysRevLett.121.251105}{\emph{Phys. Rev. Lett.} {\bfseries 121} (2018) 251105} [\href{https://arxiv.org/abs/1808.08962}{{\ttfamily 1808.08962}}].

\bibitem{Bhattacharyya:2025slf}
A.~Bhattacharyya, S.~Ghosh, N.~Kumar, S.~Kumar and S.~Pal, \emph{Love beyond {{Einstein}}: {{Metric}} reconstruction and {{Love}} number in quadratic gravity using {{WEFT}}}, \href{https://doi.org/10.48550/arXiv.2508.02785}{\emph{arXiv:2508.02785 [hep-th]} (2025) } [\href{https://arxiv.org/abs/2508.02785}{{\ttfamily 2508.02785}}].

\bibitem{Barbosa:2025uau}
S.~Barbosa, P.~Brax, S.~Fichet and L.~de~Souza, \emph{Running {{Love Numbers}} and the {{Effective Field Theory}} of {{ Gravity}}},  July, 2025.
\newblock 10.1088/1475-7516/2025/07/071.

\bibitem{Gurlebeck:2015xpa}
N.~Gürlebeck, \emph{No-hair theorem for {{Black Holes}} in {{Astrophysical Environments }}}, \href{https://doi.org/10.1103/PhysRevLett.114.151102}{\emph{Phys. Rev. Lett.} {\bfseries 114} (2015) 151102} [\href{https://arxiv.org/abs/1503.03240}{{\ttfamily 1503.03240}}].

\bibitem{Nair:2022xfm}
S.~Nair, S.~Chakraborty and S.~Sarkar, \emph{Dynamical {{Love}} for area quantized black holes}, \href{https://doi.org/10.1103/PhysRevD.107.124041}{\emph{Phys. Rev. D} {\bfseries 107} (2023) 124041} [\href{https://arxiv.org/abs/2208.06235}{{\ttfamily 2208.06235}}].

\bibitem{Motaharfar:2025ihv}
M.~Motaharfar and P.~Singh, \emph{Love {{Numbers}} of {{Covariant Loop Quantum Black Holes}}}, \href{https://doi.org/10.1103/13lp-hydg}{\emph{Phys. Rev. D} {\bfseries 112} (2025) 066008} [\href{https://arxiv.org/abs/2505.14784}{{\ttfamily 2505.14784}}].

\bibitem{Motaharfar:2025typ}
M.~Motaharfar and P.~Singh, \emph{Loop {{Quantum Gravitational Signatures}} via {{Love Numbers}}}, \href{https://doi.org/10.1103/PhysRevD.111.106018}{\emph{Phys. Rev. D} {\bfseries 111} (2025) 106018} [\href{https://arxiv.org/abs/2501.09151}{{\ttfamily 2501.09151}}].

\bibitem{Motaharfar:2025izo}
M.~Motaharfar and P.~Singh, \emph{Echoes of {{Love Beyond}} the {{Horizon}}: {{A Bridge}} to {{ Recovering Information}} from {{Black Holes}}}, \href{https://doi.org/10.48550/arXiv.2505.17189}{\emph{arXiv:2505.17189 [gr-qc]} (2025) } [\href{https://arxiv.org/abs/2505.17189}{{\ttfamily 2505.17189}}].

\bibitem{Liu:2025iby}
Y.~Liu and X.~Zhang, \emph{Quasinormal modes and tidal {{Love}} numbers of covariant effective quantum black holes with cosmological constant}, \href{https://doi.org/10.48550/arXiv.2509.12013}{\emph{arXiv:2509.12013 [gr-qc]} (2025) } [\href{https://arxiv.org/abs/2509.12013}{{\ttfamily 2509.12013}}].

\bibitem{Franzin:2024cah}
E.~Franzin, A.M.~Frassino and J.V.~Rocha, \emph{Tidal {{Love}} numbers of static black holes in anti-de {{Sitter}}}, \href{https://doi.org/10.1007/JHEP12(2024)224}{\emph{JHEP} {\bfseries 12} (2024) 224} [\href{https://arxiv.org/abs/2410.23545}{{\ttfamily 2410.23545}}].

\bibitem{Nair:2024mya}
S.~Nair, S.~Chakraborty and S.~Sarkar, \emph{Asymptotically de-{{Sitter}} black holes have non-zero tidal {{Love }} numbers}, \href{https://doi.org/10.1103/PhysRevD.109.064025}{\emph{Phys. Rev. D} {\bfseries 109} (2024) 064025} [\href{https://arxiv.org/abs/2401.06467}{{\ttfamily 2401.06467}}].

\bibitem{Kol:2011vg}
B.~Kol and M.~Smolkin, \emph{Black hole stereotyping: {{Induced}} gravito-static polarization}, \href{https://doi.org/10.1007/JHEP02(2012)010}{\emph{JHEP} {\bfseries 02} (2012) 010} [\href{https://arxiv.org/abs/1110.3764}{{\ttfamily 1110.3764}}].

\bibitem{Hui:2020xxx}
L.~Hui, A.~Joyce, R.~Penco, L.~Santoni and A.R.~Solomon, \emph{Static response and {{Love}} numbers of {{Schwarzschild}} black holes}, \href{https://doi.org/10.1088/1475-7516/2021/04/052}{\emph{JCAP} {\bfseries 04} (2021) 052} [\href{https://arxiv.org/abs/2010.00593}{{\ttfamily 2010.00593}}].

\bibitem{Rodriguez:2023xjd}
M.J.~Rodriguez, L.~Santoni, A.R.~Solomon and L.F.~Temoche, \emph{Love {{Numbers}} for {{Rotating Black Holes}} in {{Higher Dimensions}}}, \href{https://doi.org/10.1103/PhysRevD.108.084011}{\emph{Phys. Rev. D} {\bfseries 108} (2023) 084011} [\href{https://arxiv.org/abs/2304.03743}{{\ttfamily 2304.03743}}].

\bibitem{Chakravarti:2019aup}
K.~Chakravarti, S.~Chakraborty, K.S.~Phukon, S.~Bose and S.~SenGupta, \emph{Constraining extra-spatial dimensions with observations of {{ GW170817}}}, \href{https://doi.org/10.1088/1361-6382/ab8355}{\emph{Class. Quant. Grav.} {\bfseries 37} (2020) 105004} [\href{https://arxiv.org/abs/1903.10159}{{\ttfamily 1903.10159}}].

\bibitem{Cardoso:2019vof}
V.~Cardoso, L.~Gualtieri and C.J.~Moore, \emph{Gravitational waves and higher dimensions: {{Love}} numbers and {{ Kaluza-Klein}} excitations}, \href{https://doi.org/10.1103/PhysRevD.100.124037}{\emph{Phys. Rev. D} {\bfseries 100} (2019) 124037} [\href{https://arxiv.org/abs/1910.09557}{{\ttfamily 1910.09557}}].

\bibitem{Singha:2025xah}
C.~Singha and S.~Chakraborty, \emph{Tidal deformation of black holes in {{Lovelock}} gravity}, \href{https://doi.org/10.48550/arXiv.2508.14944}{\emph{arXiv:2508.14944 [gr-qc]} (2025) } [\href{https://arxiv.org/abs/2508.14944}{{\ttfamily 2508.14944}}].

\bibitem{Ma:2024few}
L.~Ma, Z.-H.~Wu, Y.~Pang and H.~Lu, \emph{Charging the {{Love}} numbers: {{Charged}} scalar response coefficients of {{Kerr-Newman}} black holes}, \href{https://doi.org/10.1103/PhysRevD.111.044003}{\emph{Phys. Rev. D} {\bfseries 111} (2025) 044003} [\href{https://arxiv.org/abs/2408.10352}{{\ttfamily 2408.10352}}].

\bibitem{Pereniguez:2025jxq}
D.~Pereñiguez and E.~Karnickis, \emph{On the non-zero {{Love}} numbers of magnetic black holes},  Sept., 2025.
\newblock 10.48550/arXiv.2509.12418.

\bibitem{Chakraborty:2025zyb}
S.~Chakraborty, P.~Heidmann and P.~Pani, \emph{Fermionic {{Love}} of {{Black Holes}} in {{General Relativity}}}, \href{https://doi.org/10.48550/arXiv.2508.20155}{\emph{arXiv:2508.20155 [gr-qc]} (2025) } [\href{https://arxiv.org/abs/2508.20155}{{\ttfamily 2508.20155}}].

\bibitem{Pang:2025myy}
X.~Pang, Y.~Tian, H.~Zhang and Q.~Jiang, \emph{Fermionic {{Love}} number of {{Reissner-Nordström}} black holes}, \href{https://doi.org/10.48550/arXiv.2510.10036}{\emph{arXiv:2510.10036 [gr-qc]} (2025) } [\href{https://arxiv.org/abs/2510.10036}{{\ttfamily 2510.10036}}].

\bibitem{Weinberg:1995mt}
S.~Weinberg, \emph{The {{Quantum}} Theory of Fields. {{Vol}}. 1: {{Foundations}}}, Cambridge University Press (1995).

\bibitem{Lopez-Ortega:2009flo}
A.~{Lopez-Ortega}, \emph{The {{Dirac}} equation in {{D-dimensional}} spherically symmetric spacetimes}, \href{https://doi.org/10.48550/arXiv.0906.2754}{\emph{arXiv:0906.2754 [gr-qc]} (2009) } [\href{https://arxiv.org/abs/0906.2754}{{\ttfamily 0906.2754}}].

\bibitem{Liu:2019rbq}
X.~Liu, S.~Van~Vooren, H.~Zhang and Z.~Zhong, \emph{Strong cosmic censorship for the {{Dirac}} field in the higher dimensional {{Reissner-Nordstrom}}–de {{Sitter}} black hole}, \href{https://doi.org/10.1007/JHEP10(2019)186}{\emph{JHEP} {\bfseries 10} (2019) 186} [\href{https://arxiv.org/abs/1909.07904}{{\ttfamily 1909.07904}}].

\bibitem{Pang:2024tco}
X.~Pang, Q.~Jiang, Y.~Xiang and G.~Deng, \emph{The precession of particle spin in spherical symmetric spacetimes}, \href{https://doi.org/10.1140/epjc/s10052-025-13894-8}{\emph{Eur. Phys. J. C} {\bfseries 85} (2025) 193} [\href{https://arxiv.org/abs/2410.04323}{{\ttfamily 2410.04323}}].

\bibitem{Wald:1984rg}
R.M.~Wald, \emph{General {{Relativity}}}, Chicago Univ. Pr., Chicago, USA (1984).

\bibitem{Camporesi:1995fb}
R.~Camporesi and A.~Higuchi, \emph{On the eigenfunctions of the {{Dirac}} operator on spheres and real hyperbolic spaces}, \href{https://doi.org/10.1016/0393-0440(95)00042-9}{\emph{J. Geom. Phys.} {\bfseries 20} (1996) 1} [\href{https://arxiv.org/abs/gr-qc/9505009}{{\ttfamily gr-qc/9505009}}].

\bibitem{Gradshteyn:1996table}
I.S.~Gradshteyn, A.~Jeffrey and I.M.~Ryzhik, \emph{Table of {{Integrals}}, {{Series}}, and {{Products}}}, Elsevier (2015), \href{https://doi.org/10.1016/C2010-0-64839-5}{10.1016/C2010-0-64839-5}.

\bibitem{Martinez:1999qi}
C.~Martinez, C.~Teitelboim and J.~Zanelli, \emph{Charged {{Rotating Black Hole}} in {{Three Spacetime Dimensions}}}, \href{https://doi.org/10.1103/PhysRevD.61.104013}{\emph{Phys. Rev. D} {\bfseries 61} (2000) 104013} [\href{https://arxiv.org/abs/hep-th/9912259}{{\ttfamily hep-th/9912259}}].

\bibitem{Bhatt:2024mvr}
R.P.~Bhatt and C.~Singha, \emph{Scalar tidal response of a rotating {{BTZ}} black hole}, \href{https://doi.org/10.1007/JHEP11(2024)154}{\emph{JHEP} {\bfseries 11} (2024) 154} [\href{https://arxiv.org/abs/2407.09470}{{\ttfamily 2407.09470}}].

\bibitem{Goldberg:1966uu}
J.N.~Goldberg, A.J.~Macfarlane, E.T.~Newman, F.~Rohrlich and E.C.G.~Sudarshan, \emph{Spin‐$s$ spherical harmonics and $\eth$}, \href{https://doi.org/10.1063/1.1705135}{\emph{J. Math. Phys.} {\bfseries 8} (1967) 2155}.

\bibitem{Penrose:1985bww}
R.~Penrose and W.~Rindler, \emph{Spinors and {{Space-Time}}: {{Volume}} 1: {{Two-Spinor Calculus}} and {{Relativistic Fields}}}, vol.~1 of \emph{Cambridge {{Monographs}} on {{Mathematical Physics}}}, Cambridge University Press, Cambridge (1984), \href{https://doi.org/10.1017/CBO9780511564048}{10.1017/CBO9780511564048}.

\end{thebibliography}
